\documentclass[12pt,a4paper]{article}

\usepackage[utf8]{inputenc}
\usepackage[T1]{fontenc}
\usepackage{amsmath,amssymb,amsfonts,array}
\usepackage{booktabs}
\usepackage[authoryear]{natbib}
\RequirePackage[dvips]{graphicx}
\usepackage{lscape,graphicx,subfigure, color}
\usepackage{epsfig}

\begin{document}

\title{BDSAR: a new package on Bregman divergence for Bayesian simultaneous autoregressive models}

\author{Ian M Danilevicz, Ricardo S Ehlers}

\date{}

\maketitle

\begin{abstract}
  
{\bf BDSAR} is an {\tt R} package (\cite{r10}) which estimates
distances between probability distributions 
and facilitates a dynamic and powerful 
analysis of diagnostics for Bayesian models from the class of
Simultaneous Autoregressive (SAR) spatial models. The package offers a new and fine
plot to compare models as well as it works in an intuitive way to
allow any analyst to easily build fine plots. These are helpful
to promote insights about influential observations in the data. 

\end{abstract}

\section{Introduction}

Spatial statistics methods were proposed to enable fitting spatially
correlated data, 
initially with Conditionally Autoregressive (CAR) models (see \cite{bes74})
and Simultaneous Autoregressive (SAR) models (see \cite{anselin88}
and \cite{Cressie}). Both models are extremely useful to understand
data from many fields as diverse as Economy, Agriculture and Oceanography. However,
the Simultaneous Autoregressive model is more parsimonious than its
``rival'' Conditionally Autoregressive model. 

Estimation procedures were developed in both frequentist
(e.g. \cite{lee96}) and Bayesian literature (e.g. \cite{oliveira2008}).
All the algorithms proposed in this paper
follow the Bayesian framework. The proposed SAR model with covariates
is presented below,
\begin{equation}
\mathbf{y} = \rho W \mathbf{y} + X \boldsymbol{\beta} + \boldsymbol{\epsilon},
\end{equation}

\noindent where $\mathbf{y} = (y_1 \ y_2 \dots y_n)'$ is an $n$ vector
of the outcomes, $X$ denotes an $ n \times k$ 
matrix of covariates, $\boldsymbol{\beta} =
(\beta_1 \ \beta_2 \dots \beta_k)'$ is a $k$ vector of linear
regression coefficients, $\boldsymbol{\epsilon} =
(\epsilon_1 \ \epsilon_2 \dots \epsilon_n)'$ is an $n$ vector of
errors and $\rho$ is the coefficient of spatial effects on $\mathbf{y}$.
Also, $W$ represents an $n\times n$ spatial weights matrix of known
constants with a zero diagonal and each row summing to one,
i.e. $w_{i,j} \in (0,1)$ with $\sum_{j=1}^{n}w_{i,j}=1$. 
As for the errors, we initially assume they are independently normally
distributed with mean zero and common variance $\sigma^2$. This is
the homoscedastic case where $\boldsymbol{\epsilon}\sim N(0,\sigma^2I_n)$ 
and $I_n$ is the identity matrix of dimension $n$. 
By using the properties of the multivariate normal distribution, it
follows that the likelihood function is given by,
\begin{equation*}\label{like}
L(\mathbf{y}|\boldsymbol{\beta}, \rho, \sigma^2) = 
\dfrac{1}{(2 \pi\sigma^2)^{n/2}} \exp \left\{ - \dfrac{1}{2 \sigma^2} (\mathbf{y} -
X \boldsymbol{\beta} - \rho W \mathbf{y})'(\mathbf{y} - X
\boldsymbol{\beta} - \rho W \mathbf{y})\right\}. 
\end{equation*}

To proceed with Bayesian inference we need to define prior
distributions for the parameters, which are assumed to be independent
{\it a priori}. The following distributions are assigned to each parameter:
$\rho \sim U(-1,1)$, $\boldsymbol{\beta} \sim N(0,\eta I_k)$ and
$\sigma^2 \sim IG(a,b)$. The Uniform between -1 and 1 takes on all
possible values for a correlation, consequently is a natural prior to $\rho$. 
$I_k$ is the identity matrix of dimension $k$, so a relatively
large value $\eta$ leads to a relatively vague prior distribution for
$\boldsymbol{\beta}$. Finally 
the prior of $\sigma^2$ is an Inverse-Gamma with hyperparameters $a$
and $b$, so that small values of $a=b$ would imply a very dispersed density.
Thus for practical purposes $a=b = 0.01$ corresponds to a
very uninformative prior. 
After defining the priors we use the Bayes theorem to determine the
joint posterior density, i.e.
\begin{eqnarray*}\label{jpos}
p(\rho, \boldsymbol{\beta},  \sigma^2 | \mathbf{y}) 
&\propto& \dfrac{1}{(\sigma^2)^{(a+1) + n/2}} \times\\
&&
\exp \left\{
- \dfrac{1}{2 \sigma^2} (\mathbf{y} - X \boldsymbol{\beta} - \rho
W \mathbf{y})'(\mathbf{y} - X \boldsymbol{\beta} - \rho W \mathbf{y})
- \dfrac{1}{2 \eta} \boldsymbol{\beta}'\boldsymbol{\beta}
-\dfrac{2b}{2\sigma^2}\right\}. 
\end{eqnarray*}

All the results in this paper correspond to the above posterior. The package provides
tools for parameter estimation, model comparison and assessment of
influential observations.
The rest of this article is organized as follows.
In Section 2,
we show how to create artificial data with a spatial pattern and how to estimate a SAR model 
using the {\bf BDSAR} package. Section 3 is dedicated to model
comparison in terms of information criteria. Section 4 concisely introduces the Bregman divergence
and clever ways to simulate it as proposed by \cite{goh} for the case of
independent observations. We then extend their methods for spatially correlated data.

\section{Create data and estimate SAR model}

The first step is to install package {\bf BDSAR} with dependencies and load it. 
Whether you do not have real data to explore, just create your own
simulated data with the {\tt sim.SAR} function. First create a weight
matrix W with functions {\tt build.a} and {\tt build.w}, later
create some exploratory and independent variables. For simplicity we
choose a design matrix $X$ normally distributed with default random number
generators from {\tt R} repository. Then we create a vector $y$ which follows a
SAR model with one covariate as well as a $z$ vector which is
identical to $y$ except to one contamination in position 10. 

\begin{verbatim}
library(BDSAR)
set.seed(2017)
n     = 50
nodes = sample(1:n,size=4*n,replace = TRUE)
A     = build.a(nodes,n)
W     = build.w(A)
x     = matrix(c(rnorm(n,-1,1),rnorm(n,2,1)),nrow=n,byrow=FALSE)
y     = sim.SAR(w=W,x=x[,1],param=c(0.9,1,0,0.3))
z     = y
if(y[1]>0) {z[10] = y[10] + quantile(y,0.99)}
if(y[1]<=0){z[10] = y[10] + quantile(y,0.01)}
\end{verbatim}

\begin{figure}[htbp]
  \centering
  \includegraphics{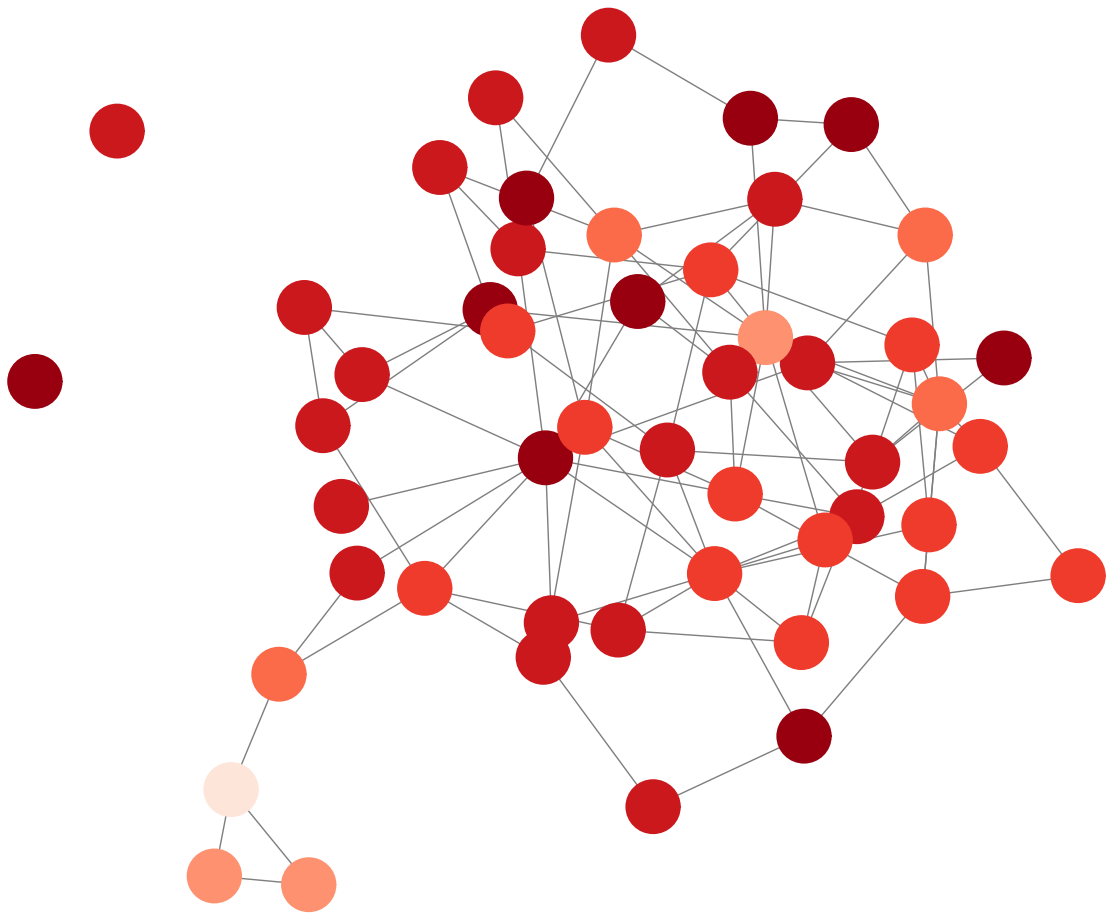}
  \caption{Graph of an adjacency matrix $A$ nodes and values of variable $y$, a simulated SAR model with one covariate}
  \label{bubles}
\end{figure}

There are many ways in which you can see the already created data.
We suggest you load four other packages which display nice
tools for elegant plots. First transform the $A$ adjacency matrix into
a network. Then define a pallete of colors and the intervals between
each color. Now you can use the function {\tt ggnet2} available
through the {\bf GGally} package to produce a fine plot
such as the one illustrated in Figure \ref{bubles}. In this figure the
collors refer to values of the variable $y$ which were grouped into classes.

\begin{verbatim}
library(network)
library(RColorBrewer)
library(classInt)
library(GGally)
net1   = network(A, directed = FALSE)
net1$y = y
colors = brewer.pal(7, "Reds")
brks   = classIntervals(y, n=7, style="equal")
brks   = brks$brks #plot the map
ggnet2(net1,node.color=colors[findInterval(net1$y, brks,all.inside=TRUE)])
\end{verbatim}

Once we have a data set and a model, we wish to estimate the model
parameters. In this paper we use Hamiltonian Monte Carlo methods (HMC, see for
example \cite{neal}), but you are free to
choose your favorite MCMC method such as Gibbs sampling. To proceed with our
example we need to load another library, the R Stan, to which our
function {\tt solv.SAR} is just a mask to help beginners. But if you
wish to design a more complex model we encourage you to write
your own codes using {\bf RStan}. Because our function was designed mainly for 
didactic objectives it is limited to just two covariates and a number of
chains between two and four. 
 
\begin{verbatim}
n_samples  = 10000
prior      = c(0.01,0.01,100)
data_cov0  = list(N=n, y=z, W=W, prior=prior)
data_cov1  = list(N=n, y=z, W=W, x=as.matrix(x[,1]), prior=prior)
data_cov12 = list(N=n, y=z, W=W, x=x, prior=prior)
m0  = solv.SAR(data=data_cov0, ncov=0, nchain=2, n_iter=n_samples)
m1  = solv.SAR(data=data_cov1, ncov=1, nchain=2, n_iter=n_samples)
m12 = solv.SAR(data=data_cov12, ncov=2, nchain=2, n_iter=n_samples)
m1
\end{verbatim}

The object $n\_samples$ is the number of samples of HMC, half of them will be discarded as burn-in. 
The vector $prior$ includes $a$, $b$ and $\sqrt{\eta}$ respectively. 
A warning here is that smaller values than 0.01 to $a$ or $b$ could produce
computational problems later at the time to estimate any divergence,
except to Kullback-Leiber divergence.  
The scalar $N$ is the length of $y$. With this information we build a
list which is a prerequisite to solve the SAR model by Stan. Furthermore
we require an additional information $ncov$ the number of covariates
in the model and $nchain$ the number of chains desired, which must
be between 2 and 4.
Enter $m1$ to display a huge table with all posterior analysis of parameters and transform parameters. 
We summarize just the information on the main parameters in Table \ref{tab1}. 

The chains for all parameters converged by $\hat{R}$ criterion (\cite{gelr92} and \cite{brooks98}). 
Yet another good result is the large values of effective sample sizes
(ESS), around five thousand of samples to all estimations, which
renders highly efficient estimates with small standard errors.
Recall that the input values to each parameter were respectively zero to
$\beta_0$, 0.3 to $beta_1$, 0.75 to $\rho$ and 1 to $\sigma$. You can
see in Table \ref{tab1} that the first two parameters were worstly
estimated than the later two. Fortunately, $\rho$ and $\sigma$ are
much more vital to the model. 

\begin{table}\centering
\caption{Summary of Posterior of SAR model parameters with one covariate.}
\vskip .2cm
\begin{tabular}{l cccccc}
\hline
parameter     & Mean   &   2.5\%  &  50\%  &   97.5\% & ESS & $\hat{R}$ \\ \hline
$\beta_0$     & -0.46  &  -1.07   & -0.46  &   0.13  & 4823   & 1 \\
$\beta_1$     & -0.39  &  -0.82   & -0.39  &   0.05  & 4862   & 1 \\
$\rho$        &  0.75  &   0.49   &  0.75  &   0.97  & 4978   & 1 \\
$\sigma$      &  2.05  &   1.39   &  1.99  &   3.02  & 5920   & 1 \\ \hline
\end{tabular}
\label{tab1}
\end{table}

\section{Comparison of models}

An important step in any data analysis is to choose between a set of
candidate models. To proceed with model comparison and selection we describe
some statistical tools in the class of information criteria. Here we
show two Bayesian criteria which generalize the well know Deviance
Information Criterion (DIC), namely the Watanabe-Akaike Information
Criterion (WAIC), designed by \cite{wata} and the leave-one-out
cross-validate (LOO-CV) (see \cite{Vehtari16}). The formulation of
each criterion is expressed in equations \ref{waic} and \ref{loo},
respectively. 

\begin{equation}\label{waic}
WAIC = -2 \sum_{i=1}^{n} \left( \log \dfrac{1}{S} \sum_{s=1}^{S}
p(y_i|\boldsymbol{\theta}^s)  \right) + 2 \sum_{i=1}^{n} \left(
Var_{s=1}^S \log  p(y_i|\boldsymbol{\theta}^s)  \right) .
\end{equation}

\begin{equation}\label{loo}
\textit{LOO-CV} = 
-2 \sum_{i=1}^{n} \left( \log \dfrac{1}{S} \sum_{s=1}^{S}  p(y_i|\boldsymbol{\theta}^{is})  \right) + 
\dfrac{2}{n} \sum_{i=1}^{n} \sum_{j=1}^{n} \left(\log\dfrac{1}{S}\sum_{s=1}^{S}p(y_j|\boldsymbol{\theta}^{is})\right).
\end{equation}

Even if the mathematical formulas are extensive, they can be easily
computed by {\bf loo} package if the log likelihood function for
transformed parameters in embeded in the Stan model.
Again this is already implemented in our {\bf BDSAR} package.

\begin{figure}[htbp]
  \centering
  \includegraphics{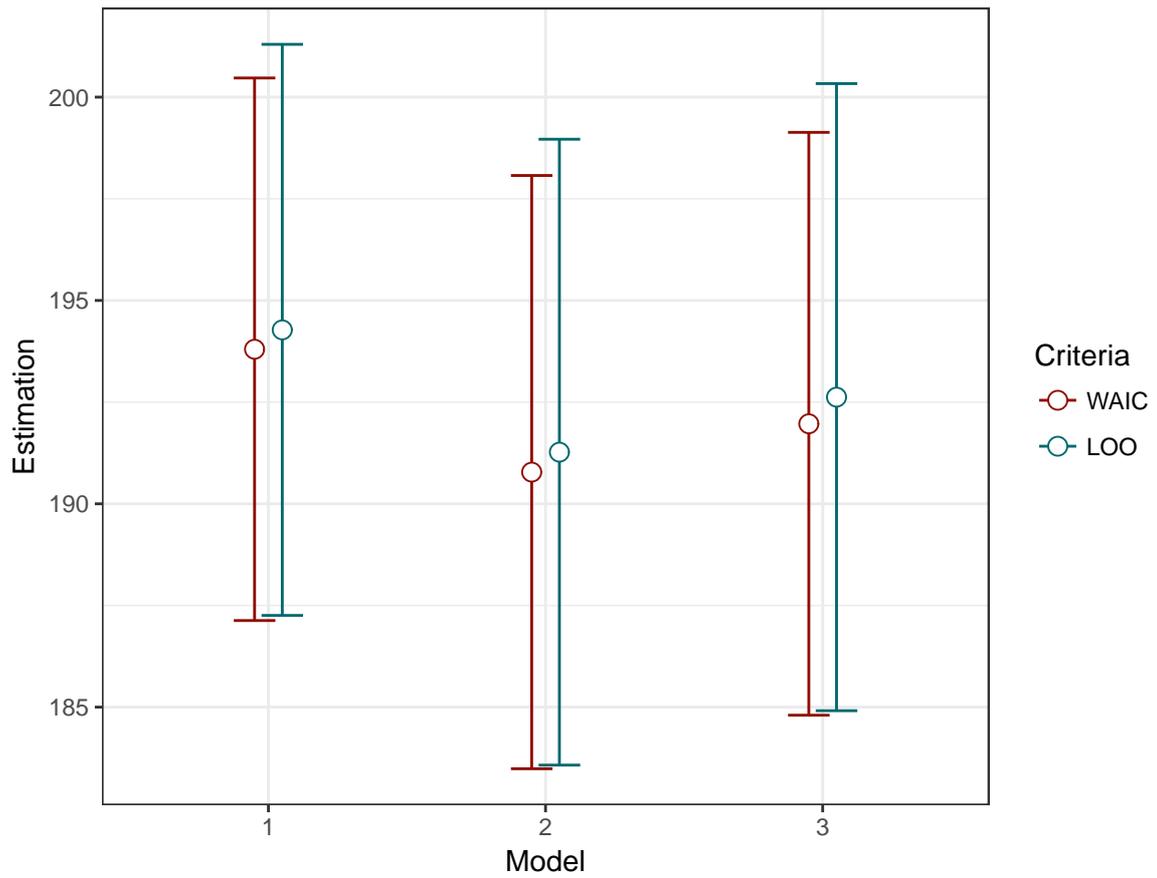}
  \caption{Comparison of three models by LOO and WAIC: M1 is a SAR
    without covariates, M2 is a SAR with one covariate, M3 is a SAR
    with two covariates.} 
  \label{3m}
\end{figure}

To be clear, the estimation of LOO-CV and WAIC is not performed by
{\bf BDSAR}, but by {\bf loo} package. 
However we offer a fine plot which summarizes the results of {\bf loo}
as you can see from the example below. 
This is the function {\tt plot.loo} which requires three components: 
\textit{n.mod} the specification of how many models are under
comparison, {\tt tab.loo} a matrix with paste results from Leave one
out, {\tt tab.waic} an analog matrix to store WAIC values. Finally you can
choose between \textit{color} equal to TRUE or FALSE if you want a
colorful plot or a black and white style. Figure \ref{3m} illustrates
the first option. We remark that this function takes advantage of 
{\bf ggplot2} package resources.

\begin{verbatim}
log_lik0  = extract_log_lik(m0, parameter_name = "log_lik")
log_lik1  = extract_log_lik(m1, parameter_name = "log_lik")
log_lik12 = extract_log_lik(m12, parameter_name = "log_lik")
loo0      = loo(log_lik0)
loo1      = loo(log_lik1)
loo12     = loo(log_lik12)
waic0     = waic(log_lik0)
waic1     = waic(log_lik1)
waic12    = waic(log_lik12)
tab.loo   = matrix(cbind(loo0,loo1,loo12))
tab.waic  = matrix(cbind(waic0,waic1,waic12))
plot.loo(n.mod=3, tab.loo, tab.waic, color=TRUE)
\end{verbatim}

We note in Figure \ref{3m} that the correct model (M2) has smaller
WAIC as well as LOO. This point estimator indicates that both criteria
work well for this kind of model, however the criteria standard errors are
relatively large and hamper stronger conclusions. Unfortunately,
this problem could not be solved by simply increasing the number of
simulated samples of HMC.

\section{Bregman Divergence}

In this section we finally discuss about the core of {\bf BDSAR}
package, i.e. about Bregman divergence which is a way to measure
distance between probability distributions. Consequently it could be
used in diagnostic analysis, e.g. checking about influential
observations. 

The reader should be familiar with the Kullback-Leiber divergence, to
which Bregman divergence is a generalization. 
We define $(X,\Omega,v)$ as a finite measure space and $f_1(x)$ and
$f_2(x)$ as two non-negative functions where any probability density
function is a special case of $f_1(x)$ or $f_2(x)$.  

Let $\psi:(0,\infty) \rightarrow  \mathbb{R}$ be a strictly convex and
differentiable function. Then the functional Bregman divergence $D_\psi$ is defined as,
 
\begin{equation}\label{bregman1}
D_\psi (f_1,f_2) = \int \{\psi[f_1(x)] - \psi[f_2(x)] - [f_1(x)-f_2(x)] \psi'[f_2(x)]     \} dv(x),
\end{equation}
 
\noindent where $\psi'$ represents the derivative of $\psi$.

The choice of the convex function $\psi$ presents some degree of
freedom. Here we follow the suggestion of \cite{goh} and restrict
attention to the class of convex functions defined by \cite{eguchi}, i.e.
$\psi_{\alpha}(x)$, with $\alpha\in\mathbb{R}$.

\begin{equation}\label{xlogx}
\psi_{\alpha}(x) = 
\left\{  
\begin{matrix}
(x^{2}-2 x + 1)/2, & \alpha=2 \\
 x\log x -x +1,   & \alpha=1 \\
-\log x + x - 1,  & \alpha = 0 \\
(x^{\alpha}-\alpha x + \alpha -1 ) / (\alpha^2 -\alpha ), & \mbox{otherwise}.
\end{matrix}
\right.
\end{equation}

Following \cite{goh}, we have a direct comparison if we take advantage
of some simulation technique, as for example Importance-Weighted
Marginal Density Estimation (IWMDE). These authors proposed this
technique to Bayesian models with independent observations, comparing
a vector $\mathbf{y}$ with $\mathbf{y}_{(i)}$, where the
second vector is equal to the first but without the $i$-th observation,  
i.e. $\mathbf{y}_{(i)} = (y_1, \dots , y_{i-1}, y_{i+1},\dots , y_n)$. 
We extend their results to correlated data using a strategy in which
our second vector incorporates an imputation of the $i$-th
observation, i.e. $\mathbf{y}_{(i)} = (y_1, \dots,\hat{y}_{i},\dots,y_n)$.

\begin{verbatim}
yhat   = y.hat(mod=m1,n=n,method=1)
draws  = as.matrix(m1)
theta  = as.matrix(cbind(draws[,3],draws[,4],draws[,1],draws[,2]))
kl     = KL.SAR(y=z,yhat=yhat,w=W,theta=theta,x=as.matrix(x[,1]),type=1)
is     = IS.SAR(y=z,yhat=yhat,w=W,theta=theta,x=as.matrix(x[,1]),prior=prior,
         dist=3,type=1)
breg   = BD.SAR(y=z,yhat=yhat,w=W,theta=theta,x=as.matrix(x[,1]),prior=prior,
         dist=3,alpha=2,type=1)
plot(kl,col=2,pch=2,ylab="D",xlab="obs")
plot(is,col=3,pch=8,ylab="D",xlab="obs")
plot(breg,col=4,pch=4,ylab="D",xlab="obs")
\end{verbatim}

Here we show how to calculate the Bregman divergence as well as two
famous specific cases: Kullback-Leiber and Itakura-Saito. First define
a vector {\tt yhat} which approximates $y$ using our
function {\tt y.hat} which requires a model, the length of $y$ and an imputation
method which could be 1 for mean or 2 for median.
The matrix $theta$ is just a compilation of parameter sampling by HMC;
again we remark that the package does not require a specific MCMC
scheme and you
could use another sampling method. However $theta$ must be organized
as $\rho$, $\sigma$, $\beta_0, \beta_1, \dots, \beta_k$ respectively.
The option $dist$ corresponds to a distribution required by the IWMDE
simulation technique. This takes values 1 for Exponential, 2 for Gamma,
3 for Normal or 4 for Multivariate Normal. According to \cite{goh} the
success of the simulation depends of a accurate choice of this
distribution, however there is no restriction to it. The parameter
$\alpha$ takes on
any real value, except 1 and 0, because this corresponds to
Kullback-Leiber and Itakura-Saito, respectively.

\begin{figure}[htbp]\centering
\includegraphics{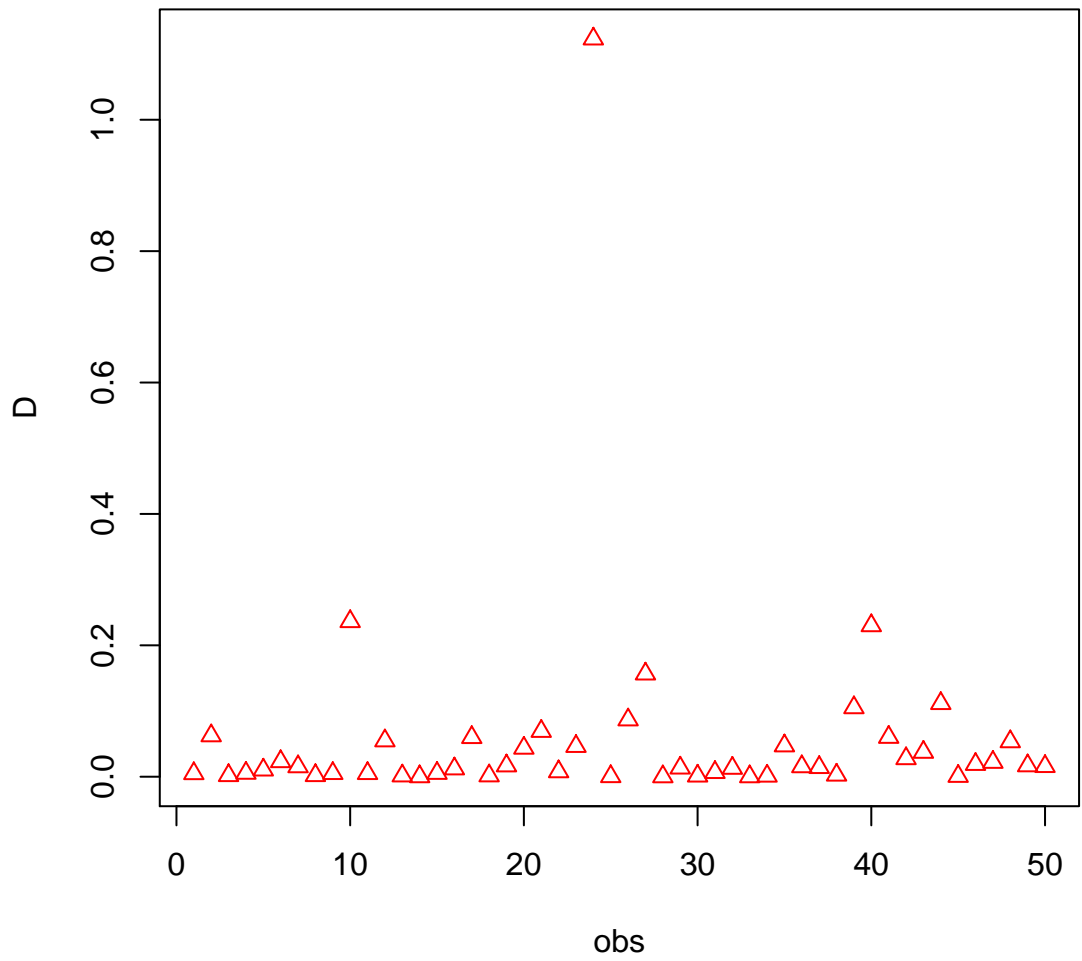}
\caption{Kullback-Leiber divergence to each observation of $z$, a SAR process with one outlier}
\label{kl}
\end{figure} 

\begin{figure}[htbp]\centering
\includegraphics{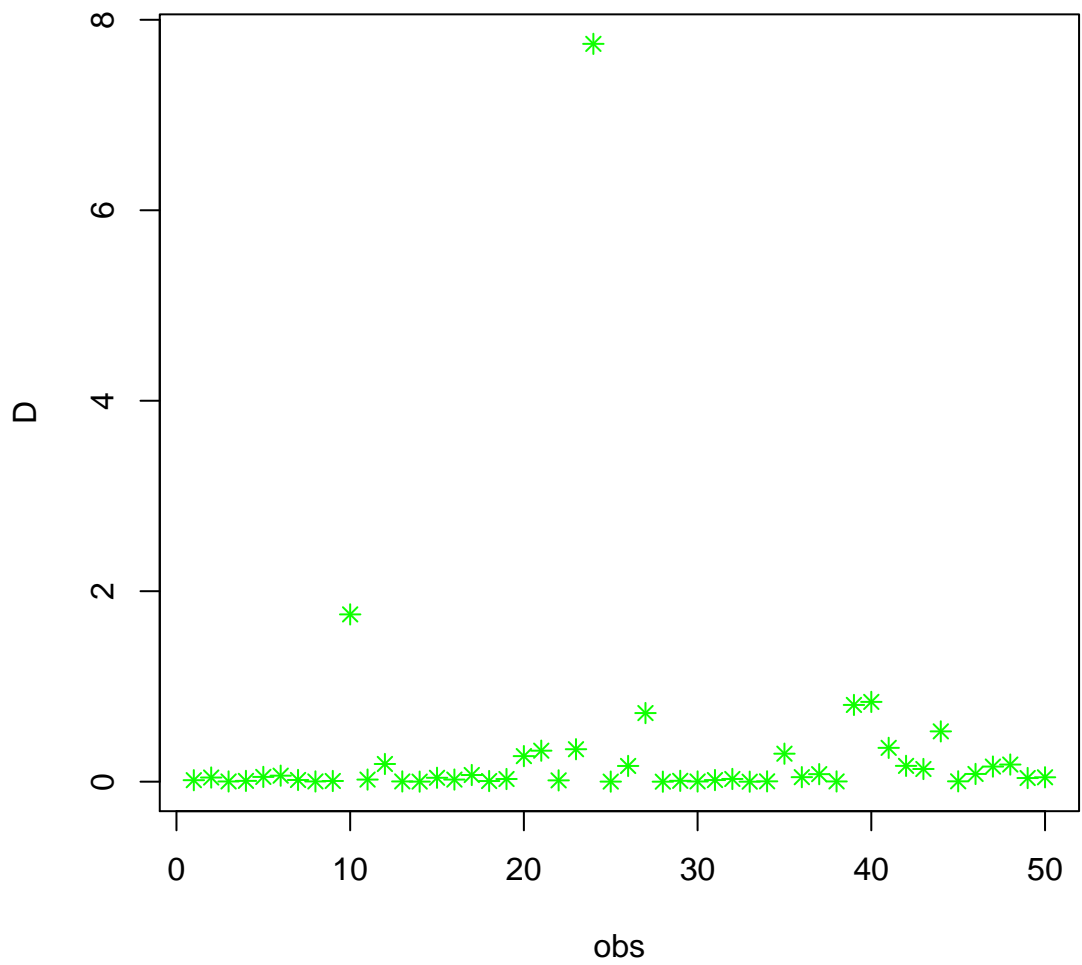}
  \caption{Itakura-Saito divergence to each observation of $z$, a SAR process with one outlier}
  \label{is}
\end{figure}

\begin{figure}[htbp]\centering
\includegraphics{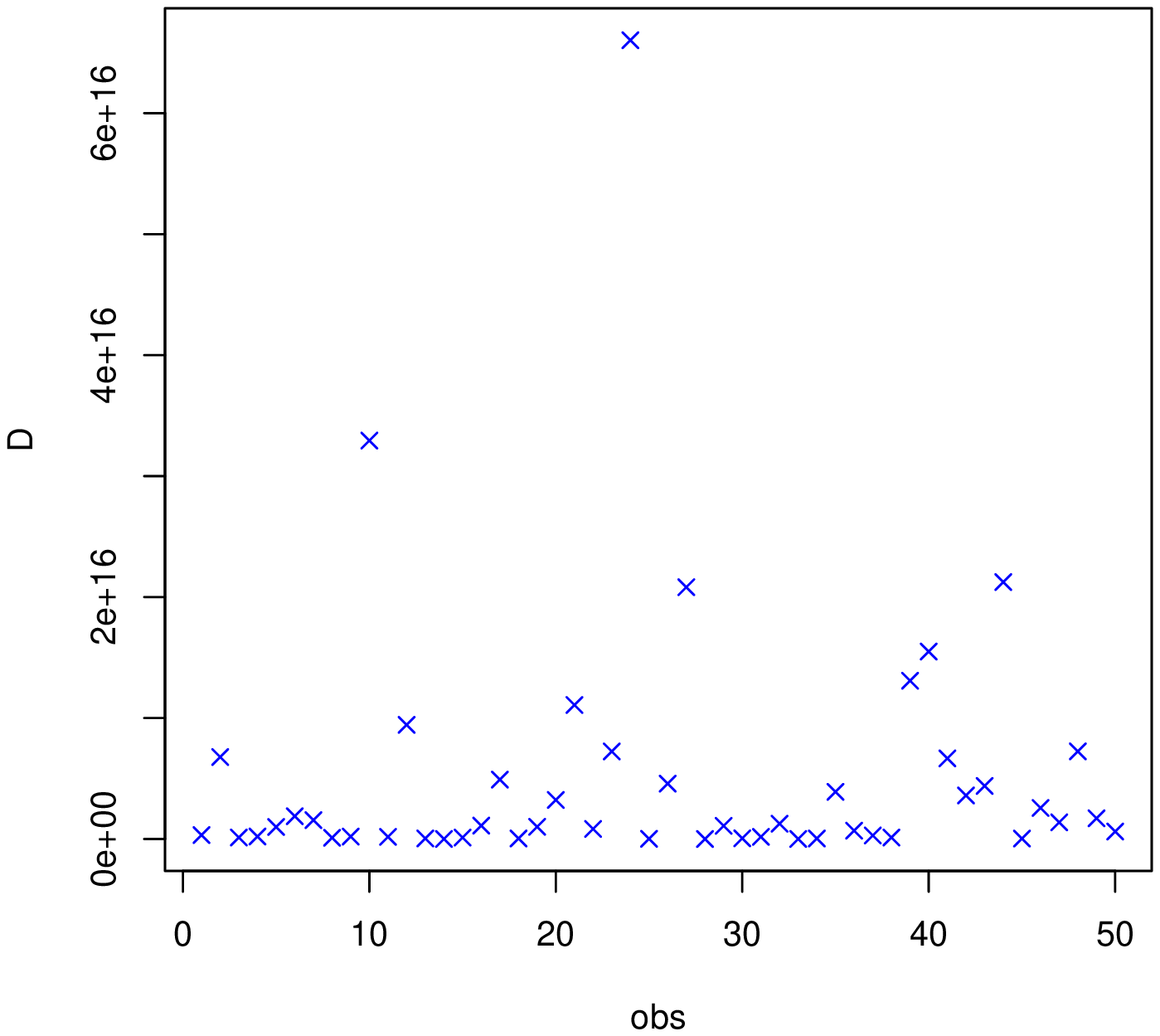}
  \caption{Bregman divergence with $\alpha=2$ to each observation of $z$, a SAR process with one outlier}
  \label{breg}
\end{figure}

In this paper we also propose an alternative way to compare the Bregman divergence between
two points, which consists of looking at a proportion, $P_i$. Once the posterior is
estimated and the matrix of simulated parameters is available it is easy to
obtain a matrix $D$ with divergences, with $s$ rows from each Monte Carlo
sampling and $i$ columns to each observation from the vector $y$. We then
count how many times each observation from $y$ displays the
supreme estimation of divergence along the MCMC iterations. Taking the
mean of these counts we obtain a
proportion, as shown in equation \ref{ppost}. 
This procedure is inspired in the work of \cite{sanb16} and is
obtained in the {\bf BDSAR} package if the option type equal to 2
is selected.

\begin{equation}\label{ppost}
P_i = \dfrac{1}{S} \sum_{s=1}^{S} \gamma_s  
\left\{
\begin{array}{l} \gamma_s = 1 \text{ , if } \hat{D}^s_i = \sup \{ \hat{D}^s\}\\
\gamma_s = 0 \text{ , otherwise. } 
\end{array}\right.
\end{equation}
\vskip .2cm

\begin{figure}[htbp]\centering
\includegraphics{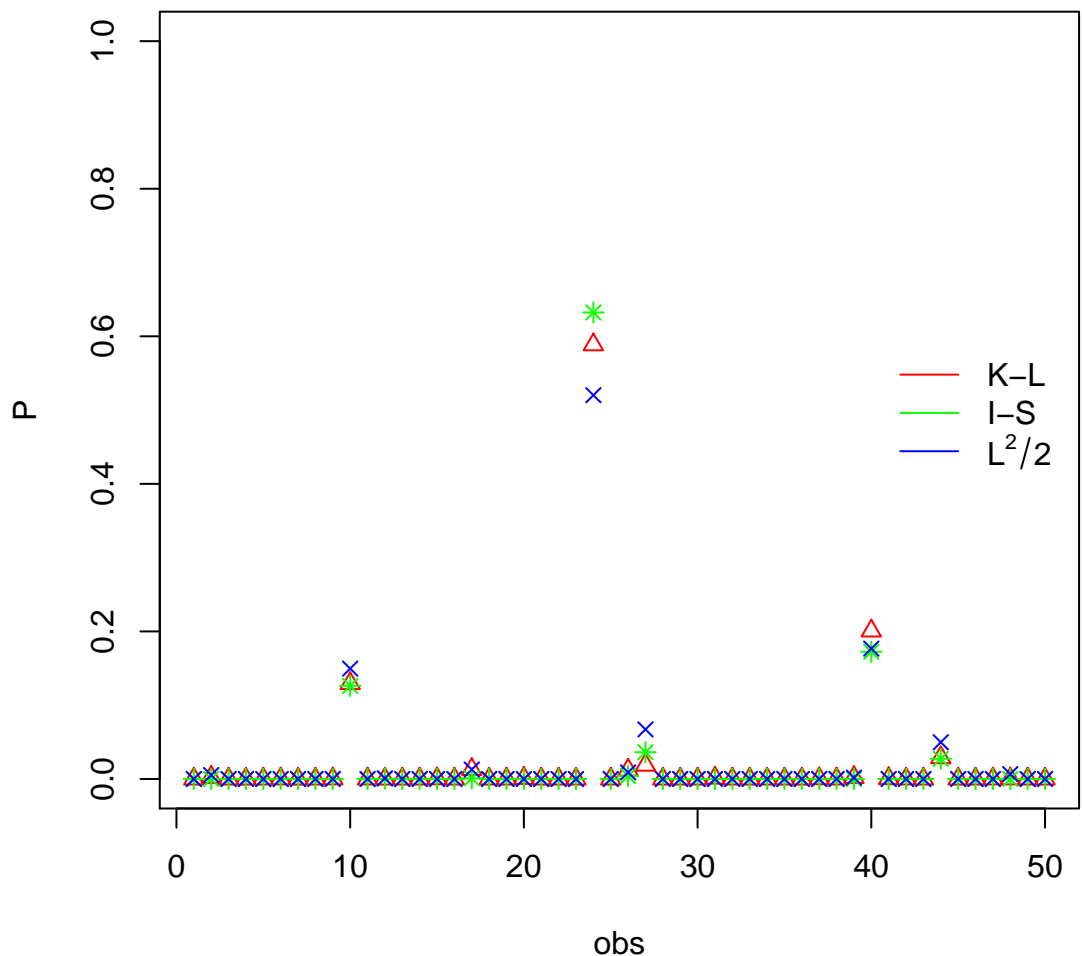}
  \caption{Comparison of three measures of divergences estimate by
    posterior probability to each observation of $z$, a SAR process
    with one outlier.} 
  \label{3d}
\end{figure}

Our last example compares three divergence measures using the type 2
option, i.e. computing the proportions. The divergences are the same
from the other example, K-L, I-S and L2/2, the Euclidean distance being a
special case of Bregman when $\alpha$ is equal to 2.
An obvious advantage of looking at the divergence as a proportion is
that it scales to a number between zero and one instead of between
zero and infinity. Consequently it becomes easier to compare
divergences as depicted in Figure \ref{3d}, where the three measures are
presented. Note that there is an outlier in position 10, but by all
divergences the position 24 is viewed as a more important influential
point. The most famous of the three is the K-L, but this measure
displays the observation 10 as the third most important, differently
of the Euclidean distance which classifies the observation 10 as the
second most relevant influence.
 
\begin{verbatim}
kl2   = KL.SAR(y=z,yhat=yhat,w=W,theta=theta,x=as.matrix(x[,1]),type=2)
is2   = IS.SAR(y=z,yhat=yhat,w=W,theta=theta,x=as.matrix(x[,1]),prior=prior,
        dist=3,type=2)
breg2 = BD.SAR(y=z,yhat=yhat,w=W,theta=theta,x=as.matrix(x[,1]),prior=prior,
        dist=3,alpha=2,type=2)
plot(c(1,n),c(0,1),ylab="P",xlab="obs",type="n")
lines(kl2,   col=2,pch=2,type="p")
lines(is2,   col=3,pch=8,type="p")
lines(breg2, col=4,pch=4,type="p")
legend("topright",col=c(2,3,4),pch=c(2,8,4),legend = c("K-L","I-S","L^2/2"))
\end{verbatim}

\section{Conclusion}

This paper offers a vignette structure which permits the reader to
follow the point with continuous code. All the examples come from
easily simulated data, so it can be reproduced by anyone.  
The sequence of arguments were organized as a full Bayesian analysis,
i.e. estimate and compare models, as well as check of assumptions
and influential observations by Bregman divergence. The {\bf BDSAR}
customize other packages to provide visual tools for analysis.  

For the analysis of influential observations we extended the
computation of Bregman divergence to spatially correlated data and
proposed a reescaling to the interval (0,1) which facilitates
comparisons. In our notation, $P_i$
corresponds to the posterior proportion of supreme cases by
observation, i.e. the most atypical observations have more chance to
present more frequently the supreme divergence.

Therefore, we believe in the relevance and usefulness of the
package {\bf BDSAR} and this paper as a useful guide to the main
ideas.

\section{Acknowledgement}

Ian Danilevicz was supported by a scholarship from the Ministry of
Education funding agency, CAPES. 
Ricardo Ehlers received support from S\~ao Paulo Research Foundation
(FAPESP) - Brazil, under grant number 2016/21137-2


\end{document}